\documentclass[paper]{JHEP3}

\usepackage{epsfig,multicol}

\newcommand\fverb{\setbox\pippobox=\hbox\bgroup\verb}
\newcommand\fverbdo{\egroup\medskip\noindent%
			\fbox{\unhbox\pippobox}\ }
\newcommand\fverbit{\egroup\item[\fbox{\unhbox\pippobox}]}
\newbox\pippobox

\def\a{\alpha}
\def\b{\beta}
\def\c{\gamma}
\def\d{\delta}
\def\e{\epsilon}

\def\l{\lambda}
\def\m{\mu}
\def\n{\nu}

\def\r{\rho}
\def\s{\sigma}

\def\w{\omega}

\def\C{\Gamma}
\def\D{\Delta}

\def\W{\Omega}

\def\pl{\partial}
\def\ul{\underline}
\def\rta{\rightarrow}

\def\Dslash{\,{\raise.15ex\hbox{/}\mkern-12mu D}}
\def\vp{v_{\rm ph}}

\title{Quantum Gravitational Optics\thanks{
Review article commissioned by `Contemporary Physics'.}}

\author{Graham M. Shore\\
	Department of Physics\\
        University of Wales, Swansea\\
        Swansea SA2 8PP, U.K.\\
	E-mail: \email{g.m.shore@swansea.ac.uk}}

\preprint{SWAT 03-375}

\abstract{
In quantum theory, the curved spacetime of Einstein's general theory of relativity
acts as a dispersive optical medium for the propagation of light. Gravitational
rainbows and birefringence replace the classical picture of light rays mapping out
the null geodesics of curved spacetime. Even more remarkably, {\it superluminal} 
propagation becomes a real possibility, raising the question of whether it is
possible to send signals into the past. In this article, we review recent
developments in the quantum theory of light propagation in general relativity
and discuss whether superluminal light is compatible with causality.
}

\begin{document}

\section{Time and Light}

Einstein's discovery of relativity nearly a century ago highlighted the fundamental nature of the
speed of light and revolutionised our concept of time. In particular, relativity predicts that
a signal moving faster than light would, from the viewpoint of some inertial observers, be moving 
backwards in time. The ideas of faster than light motion and time travel became inextricably
linked. Recently, a new perspective on these issues has arisen from the study of quantum
effects such as vacuum polarisation on the nature of light propagation in general relativity.
The classical picture of light rays mapping out the geometry of curved spacetime gives way
in quantum theory to a range of optical phenomena such as birefringence, lensing and dispersion. 
Gravitational rainbows appear. Most remarkably, even superluminal propagation appears to be possible.

In this paper, we review some of these recent developments, focusing on the issue of superluminal
light and its implications for causality. From the outset, we have to clarify what we mean by
`superluminal'. Special (general) relativity is characterised by global (local) Lorentz invariance,
that is the invariance of the laws of physics in distinct inertial frames of reference. The Lorentz
transformations introduce a new fundamental constant of nature, $c$, which in classical theory
may be identified with the unique and unambiguous speed of light. In what follows, however, we must
distinguish clearly between this fundamental constant and the actual propagation speed of light; 
when we use the word `superluminal', we simply mean `greater than $c$'. The key point is that
it is the null cones generated by the speed $c$ which divide spacetime into `timelike' and
`spacelike' regions -- a superluminal signal therefore travels into the spacelike region
and, for some observers, back in time. As we shall see, light may do precisely this. 

The propagation of light is already a rich field in classical general relativity. In the presence of 
a gravitational field, spacetime becomes curved. Light rays follow null geodesics -- {\it geodesics} 
being the analogues of shortest distant paths in curved spacetime, and {\it null} denoting that the 
frequency and wave vector satisfy the simple relation $k^2 = \w^2 - |\ul{k}|^2 = 0$ (where from now on, 
we use conventional units where $c=1$) which is equivalent to the phase velocity $v_{\rm ph} = \w/|\ul{k}|
= 1$. Light rays therefore act as tracers, mapping out the curvature of spacetime. This immediately
predicts the bending of light around a massive object, as observed by Eddington during the famous
1919 expedition to observe a solar eclipse in Brazil. Nowadays, this effect is the basis for the 
burgeoning science of gravitational lensing, which has become a crucial tool in the search for
dark matter in the universe.  

In quantum field theory, however, a whole range of new phenomena emerge. Curved spacetime 
acts as an optical medium with its own refractive index. Dispersive effects (rainbows) and
polarisation-dependent propagation (birefringence) arise. The simple identification of
the geometric null cones of curved spacetime with the physical light cones corresponding
to the actual propagation of light pulses is lost. The field of quantum gravitational optics
takes on a new richness. The big surprise, however, is the apparent prediction in some
circumstances of a superluminal speed of light.

These new developments were initiated by a pioneering paper in 1980 by 
Drummond and Hathrell \cite{DH},
who studied the effects on light propagation of vacuum polarisation (see Fig.~1) in quantum
electrodynamics. This is a uniquely quantum field theoretic process, in which a photon
metamorphoses into a virtual electron-positron pair. This gives the photon an effective
size characterised by the Compton wavelength $\l_c$ of the electron. It follows that its
propagation will be affected by gravity if the scale of the spacetime curvature is 
comparable to $\l_c$. Of course, such effects will be tiny for the curvatures associated
with normal astrophysical objects. However, this is characteristic of the physics of
quantum fields in curved spacetime, the most important example of which is the famous Hawking
\cite{Hawkrad}
radiation from black holes. Like Hawking radiation, the main importance of the effects
described here is likely to be conceptual, especially on their implications for our
understanding of the realisation of causality in theories incorporating quantum theory and gravity.  

As Drummond and Hathrell showed, the effect of vacuum polarisation is to induce direct interactions 
of $O(\a)$ between the electromagnetic field and the spacetime curvature. In other words,
the classical Maxwell equations are modified by terms explicitly involving the curvature.
This breaks the strong equivalence principle (SEP). As we explain carefully later, it is this
effective violation of the SEP which opens up the possibility that superluminal propagation
may exist without, as would be the case in special relativity, necessarily implying a
breakdown of causality. This is the key conceptual issue raised by the introduction of
quantum field theory into gravitational optics.
 
The paper is arranged as follows. In section 2, we review briefly the key points of geometric
optics, which allows us to pass from Maxwell's equations, or their quantum modifications, to
the characteristics of light propagation. Vacuum polarisation in
curved spacetime and the construction of the Drummond-Hathrell effective action is reviewed in
section 3, and its implications for photon propagation in a variety of cosmological and
black hole spacetimes are explored in section 4. Section 5 is devoted to the conceptual issues
surrounding the compatibility of superluminal light with causality. Section 6 focuses on
the crucial question of dispersion, explaining why the high-frequency limit is essential
for causality and describing recent developments 
\cite{Sfive,Ssix,Seight} in the quantum theory, including a
generalised effective action aimed at clarifying the nature of high-frequency propagation.
Finally, in section 7, we comment briefly on a number of speculative ideas in the current
literature which reflect in some way on the physics reviewed here.

\section{Geometric Optics in Curved Spacetime}

Classically, the propagation of light in general relativity is governed by Maxwell's 
equation for the electromagnetic field,
\begin{equation}
D_\m F^{\m\n} = 0
\label{eq:ba}
\end{equation}
together with the Bianchi identity
\begin{equation}
D_{[\l}F_{\m\n]} = 0
\label{eq:bb}
\end{equation}
which ensures $F_{\m\n} = \pl_\m A_\n -\pl_\n A_\m$.
The simplest way to deduce the properties of light cones and light rays 
from these equations is to use geometric optics. (For a clear introduction, see
for example ref.\cite{SEF}.) This starts from the ansatz
\begin{equation}
(A_\m + i\e B_\m + \ldots ) \exp\bigl(i {\vartheta\over\e}\bigr)
\label{eq:bc}
\end{equation}
in which the electromagnetic field is written as a slowly-varying amplitude and
a rapidly-varying phase. The parameter $\e$ is introduced as a device to keep
track of the relative order of magnitude of terms, and the Bianchi and 
Maxwell equations are solved order-by-order in $\e$.
The wave vector is identified as the gradient of the phase, 
$k_\m = \pl_\m\vartheta$. 
We also write $A_\m = A a_\m$, where $A$ represents
the amplitude itself while $a_\m$ specifies the polarisation, which
satisfies $k_\m a^\m = 0$. 

Solving the Maxwell equation, we find at $O(1/\e)$,
\begin{equation}
k^2 = 0
\label{eq:bd}
\end{equation}
while at $O(1)$,
\begin{equation}
k^\m D_\m a^\n = 0
\label{eq:be}
\end{equation}
and
\begin{equation}
k^\m D_\m(\ln A) = -{1\over2}D_\m k^\m
\label{eq:bf}
\end{equation}

Eq.(\ref{eq:bd}) shows immediately that $k_\m$ is a null vector. From its
definition as a gradient, we also see
\begin{equation}
k^\m D_\m k^\n  = k^\m D^\n k_\m = {1\over2}D^\n k^2 = 0
\label{eq:bg}
\end{equation}
Light rays, or equivalently photon trajectories, are the integral curves of 
$k^\n$, i.e.~the curves $x^\m(s)$ where $dx^\m/ds = k^\m$. These curves 
therefore satisfy
\begin{equation}
0 ~=~ k^\m D_\m k^\n 
~=~ {d^2 x^\n \over ds^2} + \C^\n_{\m\l} {dx^\m \over ds}{dx^\l \over ds}
\label{eq:bh}
\end{equation}
This is the geodesic equation. We conclude that for the usual Maxwell theory
in general relativity, light follows null geodesics. 
Eqs.(\ref{eq:bg}),(\ref{eq:be}) show that both the wave vector and the polarisation 
are parallel transported along these null geodesic rays, while Eq.(\ref{eq:bf}),
whose r.h.s.~is just (minus) the optical scalar $\theta$, shows
how the amplitude changes as the beam of rays focuses or diverges.

This picture allows us to identify the physical {\it light cones}, specified by the
dispersion relation $k^2=0$, with the geometric {\it null cones} of the background
spacetime. Moreover, light propagation is non-dispersive, i.e.~light of all
frequencies (or equivalently, photons of all momenta), travel at the same unique
speed $c$, the fundamental constant determining the local Lorentz invariance of
curved spacetime. Of course, this is why the distinction between `timelike' and
`spacelike' directions and more generally the analysis of the causal properties
of curved spacetime is traditionally expressed in the language of light signals. 
However, as we shall now see, this whole picture has to be radically revised when we
consider the quantum nature of light and its interaction with gravity.

\section{Vacuum Polarisation}

The quantum theory of photons interacting with electrons is quantum electrodynamics,
so in this section we take a first look at the effects of QED interactions on
photon propagation in curved spacetime, i.e.~in a classical background gravitational
field. 

\FIGURE
{\epsfxsize=5cm\epsfbox{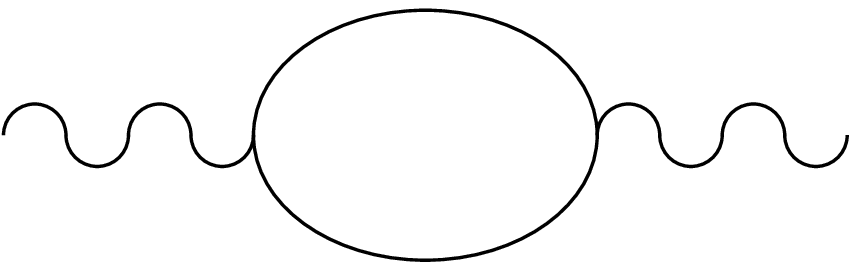}
\caption{The $O(\a)$ vacuum polarisation Feynman diagram contributing to the photon
propagator (wavy line). The solid lines represent the electron (positron) propagator 
in curved spacetime.}  \label{fig:1}}
The simplest Feynman diagram contributing to photon propagation is the so-called
`vacuum polarisation' process shown in Fig.~1. In picturesque terms, this can be
thought of as the photon splitting into a virtual electron-positron pair which
subsequently recombine. The effect is, in a certain sense, to give the photon an
effective size characterised by the Compton wavelength $\l_c = 1/m$ of the 
electron (where $m$ is its mass). It is therefore possible to think of the
photon propagating as if it were a quantum cloud of size $O(\l_c)$ rather than a
point particle. It is then plausible that if the photon passes through an anisotropic
curved spacetime whose typical curvature scale $L$ is comparable to $\l_c$, then
its motion would be affected, possibly in a polarisation-dependent way.

A rigorous analysis in QED confirms this picture. The light-cone is indeed altered
and the photon acquires a polarisation-dependent correction to its velocity 
of $O(\a \l_c^2/L^2)$. What is less predictable is that in some cases these
quantum corrections apparently produce superluminal velocities!

Technically, the consequences of vacuum polarisation are summarised in the form
of an {\it effective action}.
At one-loop order, the QED effective action is given
by
\begin{equation}
\C = \C_{\rm{M}} + \ln {\rm det} S(x,x')
\label{eq:ca}
\end{equation}
where 
\begin{equation}
\C_{\rm{M}} ~=~ -{1\over4} \int dx \sqrt{g}~F_{\m\n}F^{\m\n}
\label{eq:cb}
\end{equation}
is the free Maxwell action (which gives rise to the
standard Maxwell equation (\ref{eq:ba})) and $S(x,x')$ is the Green
function of the Dirac operator (i.e.~the electron propagator) 
in the background gravitational field, i.e.
\begin{equation}
\bigl(i\Dslash  - m\bigr) S(x,x') = {i\over\sqrt{-g}} \d(x,x')
\label{eq:cc}
\end{equation}
In fact it is more convenient to work with the differential operator
corresponding to the scalar Green function $G(x,x')$ defined by
\begin{equation}
S(x,x') = \bigl(i\Dslash  + m\bigr) G(x,x')
\label{eq:cd}
\end{equation}
so that 
\begin{equation}
\Bigl( D^2 + ie\s^{\m\n} F_{\m\n} - {1\over 4} R + m^2\Bigr) G(x,x')~~=~~
-{i\over\sqrt{-g}} \d(x,x')
\label{eq:ce}
\end{equation}
Then we evaluate $\C$ from the heat kernel, or proper time, representation
\begin{equation}
\C ~~=~~ \C_{\rm{M}} ~-~{1\over2}\int_0^\infty {ds\over s} ~e^{-i m^2 s}~ 
{\rm Tr} {\cal G}(x,x';s)
\label{eq:cf}
\end{equation}
where
\begin{equation}
{\cal D}{\cal G}(x,x';s) =  i{\pl\over\pl s}{\cal G}(x,x';s)
\label{eq:cg}
\end{equation}
with ${\cal G}(x,x';0) = \d(x,x')$. 
Here, ${\cal D}$ is the differential operator in Eq.(\ref{eq:ce}) ~at $m=0$.

A complete evaluation of this effective action, even at $O(\a)$, is impossible
with current techniques. We are forced to make two crucial approximations. The first
is a weak-field approximation for gravity -- this means we keep only terms in
the final effective action of first order in the curvature tensors $R$, $R_{\m\n}$
and $R_{\m\n\l\r}$ (the Ricci scalar, Ricci tensor and Riemann tensor respectively).
This implies our results for photon propagation are valid only to lowest order
in the parameter $\l_c^2/L^2$. The second, which we shall try to 
overcome in section 6, is a low-frequency approximation -- this means that we neglect
terms induced in the effective action which involve higher orders in derivatives
of the fields. The resulting leading-order effective action was first calculated
by Drummond and Hathrell in their pioneering paper on superluminal light \cite{DH}.
Fortunately, this already encodes much of the novel physics of photon propagation
in curved spacetime.

The Drummond-Hathrell action is
\begin{eqnarray}
\C_{\rm DH} ~~=~~ \C_{\rm M}  
+ {1\over m^2}\int dx \sqrt{-g}~&\biggl(
a R F_{\m\n}F^{\m\n} + b R_{\m\n} F^{\m\l} F^\n{}_\l
+ c R_{\m\n\l\r} F^{\m\n} F^{\l\r}\nonumber \\
{}&+ d D_\m F^{\m\l} D_\n F^\n{}_\l ~\biggr) 
\label{eq:ch}
\end{eqnarray}
where $a,b,c,d$ are perturbative coefficients of $O(\a)$, viz.
\begin{equation}
a = -{1\over144}{\a\over\pi}~~~~~~
b = {13\over360}{\a\over\pi}~~~~~~
c = -{1\over360}{\a\over\pi}~~~~~~
d = -{1\over30}{\a\over\pi}
\label{eq:ci}
\end{equation}

\section{Faster than Light?}

In this section, we show how photon propagation is modified by these new
interactions between photons and the gravitational field. As we shall see, 
this brings some surprises, notably the possibility of superluminal velocities.

\subsection{Light Cones and Birefringence}

The first step is to apply the geometric optics method described in 
section 2 to the equation of motion derived from the quantum effective 
action (\ref{eq:ch}), viz:
\begin{equation}
D_\m F^{\m\n} ~-~ {1\over m^2} \biggl[2b R_{\m\l}D^\m F^{\l\n}
+ 4c R_\m{}^\n{}_{\l\r}D^\m F^{\l\r}\biggr]~~=~~0
\label{eq:da}
\end{equation}
Here, we have neglected derivatives of the curvature tensor, which would  
be suppressed by powers of $O(\l/L)$, where $\l$ is the photon wavelength and 
$L$ is a typical curvature scale, and we have omitted the new contributions 
involving $D_\m F^{\m\n}$~: since this term is already $O(\a)$ using the
equations of motion, these contributions only affect the light cone
condition at $O(\a^2)$ and must be dropped for consistency.

Implementing the geometric optics ansatz, we quickly find the 
new light cone condition:
\begin{equation}
k^2 ~-~{2b\over m^2} R_{\m\l} k^\m k^\l ~+~{8c\over m^2} 
R_{\m\n\l\r} k^\m k^\l a^\n a^\r ~~=~~0
\label{eq:db}
\end{equation}
Since this is still homogeneous and quadratic in $k^\m$, we can write it as
\begin{equation}
{\cal G}^{\m\n}k_\m k_\n = 0
\label{eq:dc}
\end{equation}
defining ${\cal G}^{\m\n}$ as the appropriate function of the curvature
and polarisation. This immediately means that, at least in this approximation,
there is no dispersion.\footnote{Notice also that this is quite different from
the dispersion relation $p^2 = -m^2$ for a `conventional' tachyon. In this 
case (see section 5), the superluminal tachyon velocity is energy dependent.}

Now notice that in the discussion of the free Maxwell theory, we did not need to
distinguish between the photon momentum  $p^\m$, i.e.~the tangent vector 
to the light rays, and the wave vector $k_\m$ since they were simply related
by raising the index using the spacetime metric, $p^\m = g^{\m\n}k_\n$. 
In the modified theory, however, there is an important distinction. 
The wave vector, defined as the gradient of the phase, is a covariant vector 
or 1-form, whereas the photon momentum/tangent vector to the rays is a true 
contravariant vector. The relation is non-trivial. 
In fact, given $k_\m$, we should define the corresponding
`momentum' as 
\begin{equation}
p^\m = {\cal G}^{\m\n}k_\n
\label{eq:dd}
\end{equation}
and the light rays as curves $x^\m(s)$ where ${dx^\m\over ds} = p^\m$.
This definition of momentum satisfies 
\begin{equation}
G_{\m\n} p^\m p^\n = {\cal G}^{\m\n}k_\m k_\n = 0
\label{eq:de}
\end{equation}
where $G \equiv {\cal G}^{-1}$ defines a new {\it effective metric}
which determines the light cones mapped out by the geometric optics light rays.
(Indices are always raised or lowered using the true metric $g_{\m\n}$.)
The {\it ray velocity} $v_{\rm ray}$ corresponding to the momentum $p^\m$, 
which is the velocity with which the equal-phase surfaces advance, is given by
(defining components in an orthonormal frame)
\begin{equation}
v_{\rm ray} = {|\ul p|\over p^0} = {d |\ul x|\over dt}
\label{eq:df}
\end{equation}
along the ray. This is in general different from the {\it phase velocity}
\begin{equation}
v_{\rm ph} = {k^0\over|\ul k|}
\label{eq:dg}
\end{equation}
although this discrepancy only arises when there is a difference between 
the direction of propagation along the rays and the wave 3-vector. Otherwise, 
it follows directly from \ref{eq:de} ~that $v_{\rm ray}$ and $\vp$ are identical.

This shows that, at least at this level of approximation, photon propagation 
for QED in curved spacetime can be characterised as a {\it bimetric} 
theory -- the physical light cones are determined by the effective metric 
$G_{\m\n}$ and are distinct from the geometric null
cones which are fixed by the spacetime metric $g_{\m\n}$. 

It is instructive to rewrite the new light cone condition in terms of the
Weyl tensor (the trace-free part of the Riemann tensor) and the energy-momentum
tensor, using the Einstein field equation
\begin{equation}
R_{\m\n} - {1\over2}Rg_{\m\n} = 8\pi T_{\m\n}
\label{eq:dh}
\end{equation}
This gives \cite{Sthree}:
\begin{equation}
k^2 ~-~{8\pi\over m^2}(2b+4c) T_{\m\l} k^\m k^\l ~+~{8c\over m^2} 
C_{\m\n\l\r} k^\m k^\l a^\n a^\r ~~=~~0
\label{eq:di}
\end{equation}
which splits the light-cone corrections into two pieces, the first representing the
effect of matter and the second the gravitational field. The matter contribution
turns out to be relatively universal
\cite{Sthree,Scharn,Barton,LPT,Gies}; the same formula accommodates the effects
on photon propagation due to background magnetic fields, Casimir cavities, finite 
temperature environments, etc. It involves a projection of the energy-momentum
tensor whose sign is fixed by the weak-energy condition, and in all non-gravitational
situations this correction leads to a timelike $p^2$, i.e.~a reduction in the
speed of light. The second contribution, however, is unique to gravity. 

It is immediately clear that the contribution of the Weyl tensor to the light-cone
shift depends on the photon polarisation $a_\m$. This interaction therefore produces 
a polarisation-dependent shift in the velocity of light -- 
{\it gravitational birefringence}.
Moreover, it has a special property which is not immediately evident but can be
proved simply using symmetry properties of the Weyl tensor (see \cite{Sthree}). The two
physical transverse polarisations give a contribution of the same magnitude but
of {\it opposite} sign. This has a striking consequence: if we consider Ricci-flat 
spacetimes, so that the Weyl interaction is the only contribution to the light-cone shift,
it follows that if one polarisation produces a timelike shift in $p^2$, the other must
necessarily produce a spacelike shift. That is, the light cone (\ref{eq:di}) necessarily
predicts {\it superluminal} velocities!

\subsection{Cosmological and Black Hole Spacetimes}

To understand these new phenomena in more detail, we now look at two specific
examples. First, we consider the Friedmann-Robertson-Walker (FRW) spacetime which 
describes standard big-bang cosmology -- this is Weyl-flat so only the 
energy-momentum term in Eq.(\ref{eq:di}) contributes to the velocity shift.
Second, we consider a Ricci flat case, where only the Weyl term contributes,
viz.~the Schwarzschild spacetime describing a non-rotating black hole.

The FRW metric is
\begin{equation}
ds^2 = dt^2 - R^2(t)\Bigl({dr^2\over 1-kr^2} + r^2(d\theta^2 + \sin^2\theta d\phi^2)\Bigr)
\label{eq:dj}
\end{equation}
where $R(t)$ is the scale factor and $k =0,\pm 1$ determines whether the 3-space is flat,
open or closed. The energy-momentum tensor is
\begin{equation}
T_{\m\n} = (\r + P)n_\m n_\n - P g_{\m\n}
\label{eq:dk}
\end{equation}
with $n^\m$ specifying the time direction in a comoving
orthonormal frame. $\r$ is the energy density and $P$ is the
pressure, which in a radiation-dominated era are related by $\r - 3P = 0$.

Since the FRW metric is Weyl flat, the light cone (\ref{eq:di}) is simply
\begin{equation}
k^2 = \zeta~ T_{\m\n} k^\m k^\n 
\label{eq:dl}
\end{equation}
where $\zeta = {8\pi\over m^2}(2b+4c)$. 
The corresponding phase velocity
\begin{equation}
v_{\rm ph} = 1 + {1\over 2}\zeta (\r + P)
\label{eq:dll}
\end{equation}
is independent of polarisation and is found to be
superluminal \cite{DH}.

At first sight, this may look surprising given that $k^2 > 0$, its sign
fixed by the weak energy condition $T_{\m\n} k^\m k^\n \ge 0$.
However, if instead we consider the momentum along the rays,
$p^\m = {\cal G}^{\m\n}k_\n$,
we find
\begin{equation}
p^2 = g_{\m\n}p^\m p^\n = -\zeta (\r + P) (p^0)^2
\label{eq:dm}
\end{equation}
and 
\begin{equation}
v_{\rm ray} = {|\ul p|\over p^0} = 1 + {1\over2}\zeta(\r + P)
\label{eq:dn}
\end{equation}
The effective metric $G = {\cal G}^{-1}$ is (in the orthonormal frame)
\begin{equation}
G ~=~ \left(\matrix{1 + \zeta\r &0&0&0\cr
0&-(1-\zeta P)&0&0\cr 0&0&-(1-\zeta P)&0\cr 0&0&0&-(1-\zeta P)\cr}\right)
\label{eq:do}
\end{equation}
In this case, therefore, we find equal and superluminal
velocities $v_{\rm ph} = v_{\rm ray}$ and $p^2 < 0$ is manifestly
spacelike as required. 

For a second example \cite{DH, Sone, Stwo}, 
consider the Ricci-flat Schwarzschild spacetime, with
metric
\begin{equation}
ds^2 = \Bigl(1-{2M\over r}\Bigr)dt^2 - \Bigl(1-{2M\over r}\Bigr)^{-1} dr^2 
- r^2(d\theta^2 + \sin^2\theta d\phi^2) 
\label{eq:dp}
\end{equation}
In the classical theory, with $k^2=0$, there is a special solution of the
null geodesic equations describing a light ray in a circular orbit with
specified radius $r=3M$. Using the new light cone (\ref{eq:di}) and choosing
$k_\m$ and $a_\m$ to represent photons in a circular orbit with transverse
polarisations, we find 
\begin{equation}
k^2 = \pm {8c\over m^2} {3M\over r^3} (k^0)^2
\label{eq:dq}
\end{equation}
where $k^0$ is the time component of the wave vector in an orthonormal frame. 
The phase velocity is therefore
\begin{equation}
v_{\rm ph} = {k^0\over k^3} = 1 \pm {4c\over m^2} {3M\over r^3}
\label{eq:dr}
\end{equation}
and evaluating at $r=3M$ gives the phase velocity of the new circular orbit,
up to $O(\a)$. Crucially, we see that one of the polarisations has a 
superluminal velocity.

The radius of the circular photon orbit is changed to $r=3M + O(\a)$, and similarly
the location of the effective ergosphere for the rotating black hole (Kerr metric)
is shifted by polarisation-dependent $O(\a)$ corrections due to the modified speed of 
light \cite{Stwo}. However, the event horizon itself has a special status 
\cite{Sthree,Gibb}. It can be shown that for a normal-directed null vector $k_\m$, 
both terms in the modified light cone (\ref{eq:di}) vanish \cite{Hawk}. 
It seems that while in general the quantum-corrected
light cones differ from the geometric null cones, the geometric event horizon remains
a true horizon for the propagation of physical photons.

\subsection{Observability?}

A similar calculation in the Schwarzschild metric shows that the angle of deflection
of light around a star (one of the classic tests of general relativity) is modified.
If the classical deflection angle is $\D \phi$, then the quantum shift is given
by \cite{DH}
\begin{equation}
\d(\D\phi) = \pm {8c\over m^2} {1\over R^2} \D\phi
\label{eq:ds}
\end{equation}
where $R$ is the distance of closest approach. This gravitational birefringence
effect would also mean that there is in principle a polarisation dependence in 
{\it gravitational lensing}. Is this observable?

The problem is that the magnitude of all these quantum gravitational optics phenomena
is incredibly small for astrophysical scales. Recall that the order of magnitude
of the shift in the speed of light is $O(\a \l_c^2/L^2)$ for a typical curvature
scale $L$. It is therefore suppressed by the square of the ratio of a quantum scale
$(\l_c)$ to an astrophysical curvature scale $L$. This is tiny -- for
a solar mass black hole the correction (\ref{eq:dr}) to the speed of light is
$O\bigl(\a m_{pl}^4/(m^2 M^2)\bigr) \sim O(10^{-34})$, while the correction 
(\ref{eq:ds}) to the bending of light for solar parameters is only of $O(10^{-47})$. 
Of course this is not surprising -- it follows immediately from the discussion of the 
physics of vacuum polarisation in section 3 -- and is typical of all phenomena in the 
theory of quantum fields in curved spacetime, such as the famous example of Hawking 
radiation from black holes. The effect only becomes large for quantum-scale black holes, 
which may have arisen in the very early universe.
Like Hawking radiation, therefore, the main interest in this phenomenon of
quantum-induced superluminal propagation is probably conceptual: the existence of 
thermal Hawking radiation forces us to reconsider the consistency of quantum mechanics
with gravity, while this superluminal effect challenges our understanding of time 
and causality. 

Similarly, in the case of the FRW spacetime, the change in the speed of light only
becomes appreciable when the curvature becomes of order the quantum scale, at which 
point our weak-field approximation breaks down and we can only guess at the possible
effects. Nevertheless, in the radiation dominated era, 
where $\r(t) = {3\over 32\pi} t^{-2}$, we have
\begin{equation}
v_{\rm ph} = 1 + {1\over 16\pi}\zeta~ t^{-2}
\label{eq:dt}
\end{equation}
which, as already observed in ref.\cite{DH}, increases towards the early
universe. Although this expression is only reliable in the perturbative
weak-field regime, it is nonetheless intriguing that QED
predicts a rise in the speed of light in the early universe. It is interesting
to speculate whether this superluminal effect persists for high curvatures near 
the initial singularity and whether it could play a role in resolving
the horizon problem in cosmology.
 
Beyond these potential cosmological or astrophysical effects, there is a
potential obstruction in principle to the observation of these superluminal
phenomena. The argument runs as follows \cite{DH}. If we assume that a typical
time over which propagation can be followed is characterised by the 
curvature scale $L$, then the length difference between 
paths corresponding to different polarisations is $~ \a \l_c^2/L$. Very loosely,
let us say that observability requires this to be $O(\l)$, where $\l$ is the
wavelength of the light. The condition for observability is
therefore~ $\a \l_c^2/(\l L) > 1$. But the derivation required not only the 
weak-field assumption $L\gg\l_c$ but also assumed (because of the truncated
derivative expansion in the action) that the photon frequency is small, i.e.
$\l \gg \l_c$. Now in practice, spectroscopic techniques allow very much 
more sensitive detection so the $O(1)$ on the right hand side of the observability
criterion could be many orders of magnitude less (indeed factors
of $~ 10^{10}$ are achieved in the interferometry used in gravitational 
wave detectors). Nevertheless, this does mean
that to ensure that the superluminal phenomenon is genuinely observable as a 
matter of principle and that there is no absolute obstruction to its measurement, we
need to extend the derivation of the light cone condition beyond the low-frequency
approximation. This is addressed in section 6, where we study the whole question
of dispersion.

\section{Causality}

The ability to send signals `faster than light' is inextricably linked in
most people's minds with the possibility of time travel. However, this
identification is founded mainly on intuition arising from special relativity.
As we shall see, the correspondence between superluminal propagation and
causality in general relativity is rather more subtle. 
 
\subsection{Superluminal Propagation in Special and General Relativity}

We therefore begin by considering superluminal propagation in special relativity.
The first important observation is that given a superluminal signal we can
always find a reference frame in which it is travelling backwards in time.
This is illustrated in Fig.~2. 

\FIGURE
{\epsfxsize=11cm\epsfbox{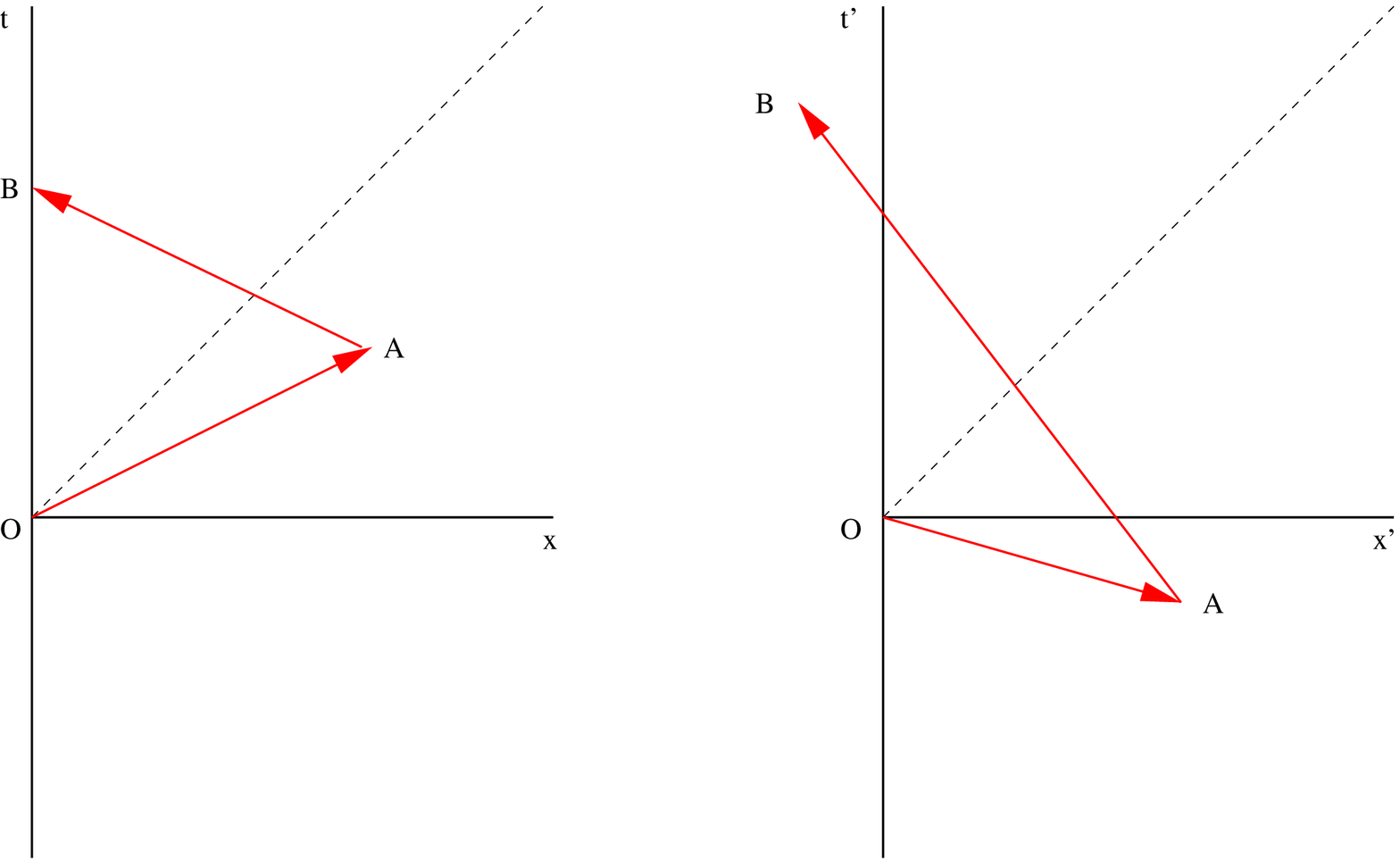}
\caption{A superluminal $(v>1)$ signal OA which is forwards 
in time in frame ${\cal S}$ is backwards in time in a frame ${\cal S}'$ moving
relative to ${\cal S}$ with speed $u>{1\over v}$. However, the return path with
the same speed in ${\cal S}$ arrives at B in the future light cone of O, 
independent of the frame.}  \label{fig:2}}

Suppose we send a signal from O to A at 
speed $v>1$ (in $c=1$ units) in frame ${\cal S}$ with coordinates $(t,x)$.
In a frame ${\cal S}'$ moving with respect to ${\cal S}$ with velocity 
$u>{1\over v}$, the signal travels backwards in $t'$ time, as follows
immediately from the Lorentz transformation. To see this in detail, recall that from
the Lorentz transformations we have 
$t'_A = \c(u)t_A (1-uv)$ 
and 
$x'_A = \c(u)x_A (1-{u\over v})$
For the situation realised in Fig.~2, we require both $x'_A >0$ and $t'_A <0$,
that is ${1\over v} < u < v$, 
which admits a solution only if $v>1$.

The important point for our considerations is that this {\it by itself} 
does not necessarily imply a violation of causality. For this, we require that 
the signal can be returned from A to a point in the past light cone of O. 
However, if we return the signal from A to B with the same speed in frame ${\cal S}$,
then of course it arrives at B in the future cone of O. The situation is physically
equivalent in the Lorentz boosted frame ${\cal S}'$ -- the return signal travels 
forward in $t'$ time and arrives at B in the future cone of O. This, unlike
the assignment of spacetime coordinates, is a frame-independent statement.

\FIGURE
{\epsfxsize=5cm\epsfbox{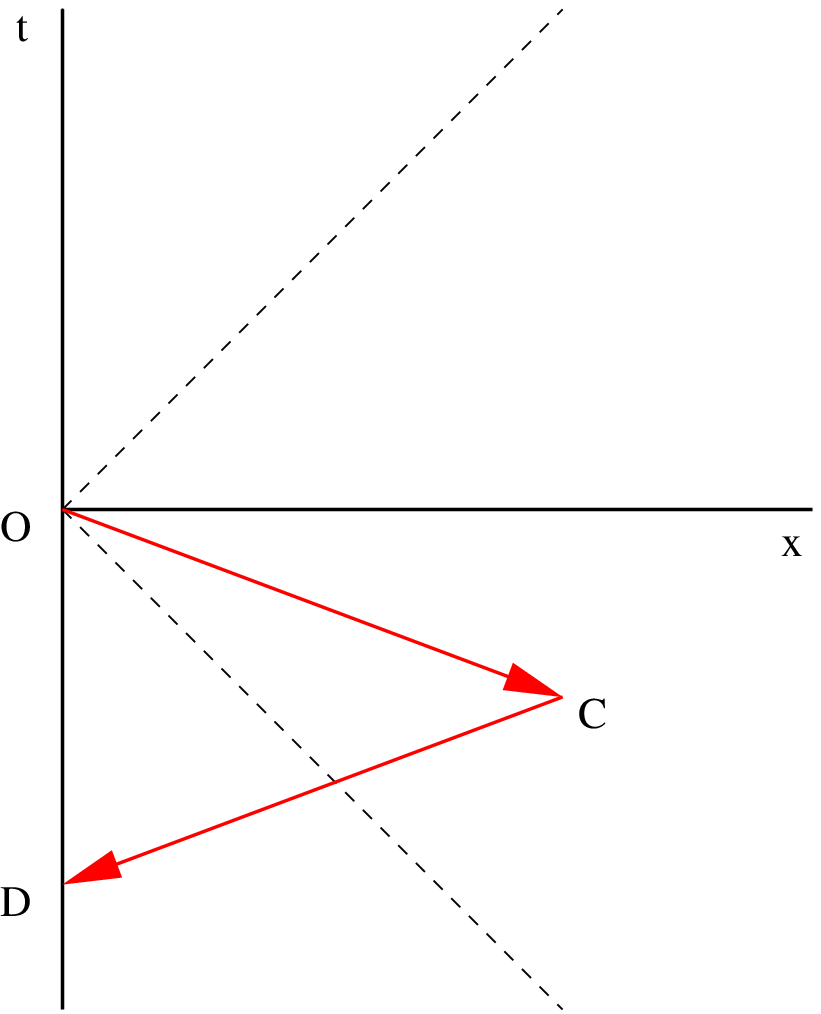}
\caption{A superluminal $(v>1)$ signal OC which is backwards 
in time in frame ${\cal S}$ is returned at the same speed to point D in the past light
cone of O, creating a closed time loop.}  \label{fig:3}}
The problem with causality arises from the scenario illustrated in Fig.~3.
Clearly, if a backwards-in-time signal OC is possible in frame ${\cal S}$,
then a return signal sent with the same speed will arrive at D in the past light
cone of O creating a closed time loop OCDO.
The crucial point is that local Lorentz invariance of the laws 
of motion implies that if a superluminal signal such as OA is possible, then so is one
of type OC \footnote{For example, suppose we have a conventional tachyon with a
spacelike momentum $p^2 = -m^2 < 0$. Setting $p^\m = {dx^\m/ ds}$ so that its 
velocity is $v_{tach} = {d|\ul{x}|/dt} = {|\ul{p}|/ p^0}$, we find
$v_{tach} = \sqrt{1 + {m^2\over E^2}}$
where we have used the notation $E=p^0$ and as usual set $c=1$. The velocity therefore
depends on the energy (note that the dispersion relation $p^2+m^2 = 0$, while Lorentz 
invariant, is not homogeneous in $p$), so by an appropriate Lorentz transformation
we can change the superluminal tachyon velocity at will. The general result is that if 
the dynamical equations are invariant under local Lorentz transformations and 
superluminal motion of type OA (Fig.~2) is possible, then so is `backwards-in-time' 
motion of type OC (Fig.~3).}, since it is just given by an appropriate Lorentz boost.
The existence of {\it global} inertial frames then guarantees the existence of
the return signal CD (in contrast to the situation in Fig.~2 viewed in the
${\cal S}'$ frame). 

The moral is that {\it both} conditions must be met in order to guarantee the 
occurrence of unacceptable closed time loops -- the existence of a superluminal signal
{\it and} global Lorentz invariance. Of course, since global Lorentz invariance
(the existence of global inertial frames) is the essential part of the structure
of special relativity, we recover the conventional wisdom that in this theory, 
superluminal propagation is indeed in conflict with causality. 

\pagebreak
So much for special relativity. The situation is, however, crucially different in 
general relativity. This is formulated on the basis of the {\it weak} equivalence 
principle, which we understand here as the statement that a {\it local} inertial 
frame exists at each point in spacetime. This implies that spacetime in general 
relativity is a Riemannian manifold. However, local Lorentz invariance alone is not
sufficient to establish the link between superluminal propagation
and causality violation. This is usually established by adding 
a second, dynamical, assumption. The {\it strong}
equivalence principle (SEP) states that the laws of physics should be identical in
the local frames at different points in spacetime, and that they should
reduce to their special relativistic forms at the origin of each local frame.
It is the SEP which takes over the role of the existence of global inertial frames
in special relativity in establishing the incompatibility of superluminal
propagation and causality.

However, unlike the weak equivalence principle, which underpins the essential
structure of general relativity, the SEP is merely a simplifying
assumption about the dynamics of matter coupled to gravitational fields.
Mathematically, it requires that matter or electromagnetism is {\it minimally
coupled} to gravity, i.e.~with interactions depending only on the connections
but not the local curvature. This ensures that at the origin of a local frame,
where the connections may be Lorentz transformed locally to zero, the dynamical
equations recover their special relativistic form. In particular, the SEP is
violated by interactions which explicitly involve the curvature, such as
those occurring in the quantum Drummond-Hathrell action and the consequent
modified light cones.

We discuss below the question of whether this specific realisation of superluminal 
propagation is in conflict with causality, using the concept
of {\it stable causality} described in ref.\cite{HE}. Notice though that by
violating the SEP, we have evaded the {\it necessary} association of superluminal
motion with causality violation that held in special relativity. Referring back to
the figures, what is established is the existence of a signal of type OA,
which as we saw, does not by itself imply problems with causality even though frames 
such as ${\cal S}'$ exist {\it locally} with respect to which motion is backwards in 
time. However, since the SEP is broken, even if a local frame exists in which the 
signal looks like OC, it does {\it not} follow that a return path CD is allowed. 
The signal propagation is fixed, determined locally by the spacetime curvature.

\subsection{Stable Causality}

One special case where causality is realised in a particularly simple way is in
globally hyperbolic spacetimes, where the manifold admits a foliation 
into a set of spacelike Cauchy surfaces with fibres given by timelike geodesics.
It is not hard to imagine that the same structure could be preserved using the 
effective metric $G_{\m\n}$ to define `spacelike' or `timelike', especially if
$G_{\m\n}$ is only perturbatively different from the actual spacetime metric $g_{\m\n}$.
But this would be a global question and the preservation of global hyperbolicity
is not {\it a priori} guaranteed. 

The clearest criterion for causality in general involves the concept of 
{\it stable causality} discussed, for example, in the monograph of Hawking and 
Ellis \cite{HE} (see also \cite{LSV}). Proposition 6.4.9 states the required
definition and theorem:

\noindent $\bullet$ A spacetime manifold $({\cal M},g_{\m\n})$ is {\it stably causal} 
if the metric $g_{\m\n}$ has an open neighbourhood such that ${\cal M}$ has no closed 
timelike or null curves with respect to any metric belonging to that 
neighbourhood.

\noindent $\bullet$ Stable causality holds everywhere on ${\cal M}$ if and only if there
is a globally defined function $f$ whose gradient $D_\m f$ is everywhere non-zero
and timelike with respect to $g_{\m\n}$.  

According to this theorem, the absence of causality violation in the form of closed
timelike or lightlike curves is assured if we can find a globally defined function $f$ whose
gradient is timelike {\it with respect to the effective metric} $G_{\m\n}$ for light
propagation. $f$ then acts as a global time coordinate. 

We can immediately give one example where this condition does hold. This is the FRW 
spacetime discussed in section 4.2. In this case, the effective metric (in the 
unperturbed orthonormal frame) is given by (\ref{eq:do}). We simply 
use the cosmological time coordinate $t$ as the globally
defined function $f$. We need only then check that $D_\m t$ defines a 
timelike vector with respect to the effective metric $G_{\m\n}$. 
This is true provided $G_{00} > 0$, which is certainly satisfied
by Eq.(\ref{eq:do}). So at least in this case, superluminal propagation 
is compatible with causality.

\section{Gravity's Rainbow}

So far, our discussion of photon propagation has been based on the low-frequency
approximation arising from the effective action (\ref{eq:ch}). However, we know that
this cannot be the whole story. In this section, we investigate dispersion -- how
the speed of light depends on its frequency -- and explain carefully why the
high-frequency limit is critical for causality. We begin with a review of
some basic properties of light propagation in classical optics.

\subsection{The `Speeds of Light'}

An illuminating discussion of wave propagation in a simple
dispersive medium is given in the classic work by Brillouin \cite{Brill}. 
This considers propagation of a sharp-fronted pulse of waves in a medium 
with a single absorption band, with refractive index $n(\w)$:
\begin{equation}
n^2(\w) = 1 - {a^2\over \w^2 -\w_0^2 + 2i\w\r}
\label{eq:fa}
\end{equation}
where $a,\r$ are constants and $\w_0$ is the characteristic frequency
of the medium. Five\footnote{In fact, if we take into account the distinction
discussed in section 4 between the {\it phase velocity} $v_{\rm ph}$ and the 
{\it ray velocity} $v_{\rm ray}$, and include the fundamental speed of light 
constant $c$ from the Lorentz transformations, we arrive at {\it seven} 
distinct definitions of `speed of light'.}
distinct velocities are identified: the {\it phase
velocity} $v_{\rm ph} = {\w\over{|\ul k|}} = \Re{1\over n(\w)}$, 
{\it group velocity} $v_{\rm gp} = {d\w\over d|\ul k|}$,
{\it signal velocity} $v_{\rm sig}$, {\it energy-transfer velocity} 
$v_{\rm en}$ and {\it wavefront velocity} $v_{\rm wf}$, with precise 
definitions related to the behaviour of contours and saddle points in the
relevant Fourier integrals in the complex $\w$-plane.
Their frequency dependence is illustrated in Fig.~4.

\FIGURE
{\epsfxsize=10cm\epsfbox{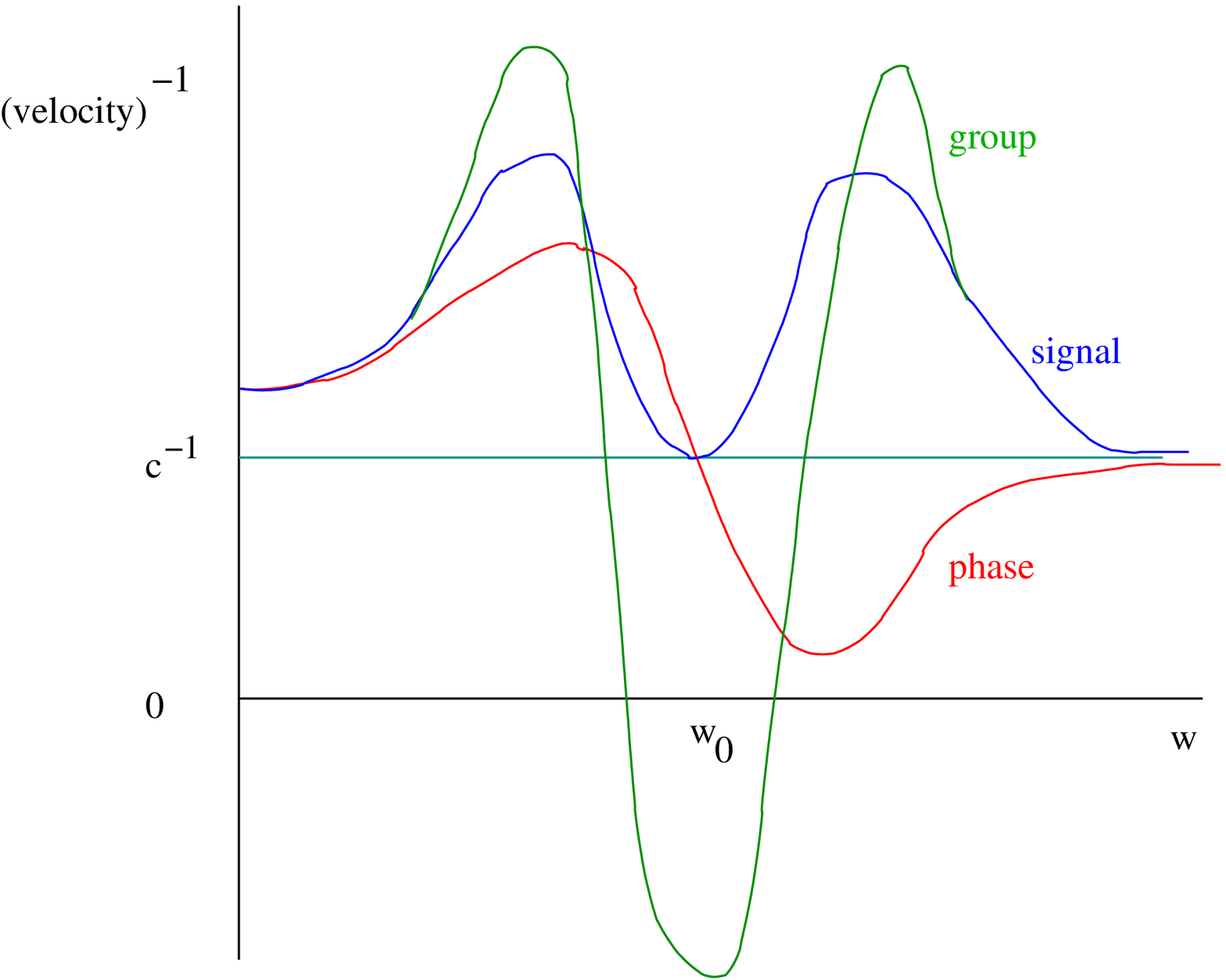}
\caption{ Sketch of the behaviour of the phase, group and 
signal velocities with frequency in the model described by the refractive
index Eq.(\ref{eq:ca}). The energy-transfer velocity (not shown) is always 
less than $c$ and becomes small near $\w_0$. The wavefront speed is 
identically equal to $c$.} \label{fig:4}}

As the pulse propagates, the first disturbances to arrive are
very small amplitude waves, `frontrunners', which define the wavefront
velocity $v_{\rm wf}$. These are followed continuously by waves with amplitudes
comparable to the initial pulse; the arrival of this part of the complete
waveform is identified in ref.\cite{Brill} as the signal velocity $v_{\rm sig}$.
As can be seen from Fig.~4, it essentially coincides with the more familiar
group velocity for frequencies far from $\w_0$, but gives a much more
intuitively reasonable sense of the propagation of a signal than the group 
velocity, whose behaviour in the vicinity of an absorption band is
relatively eccentric.

\FIGURE
{\epsfxsize=8cm\epsfbox{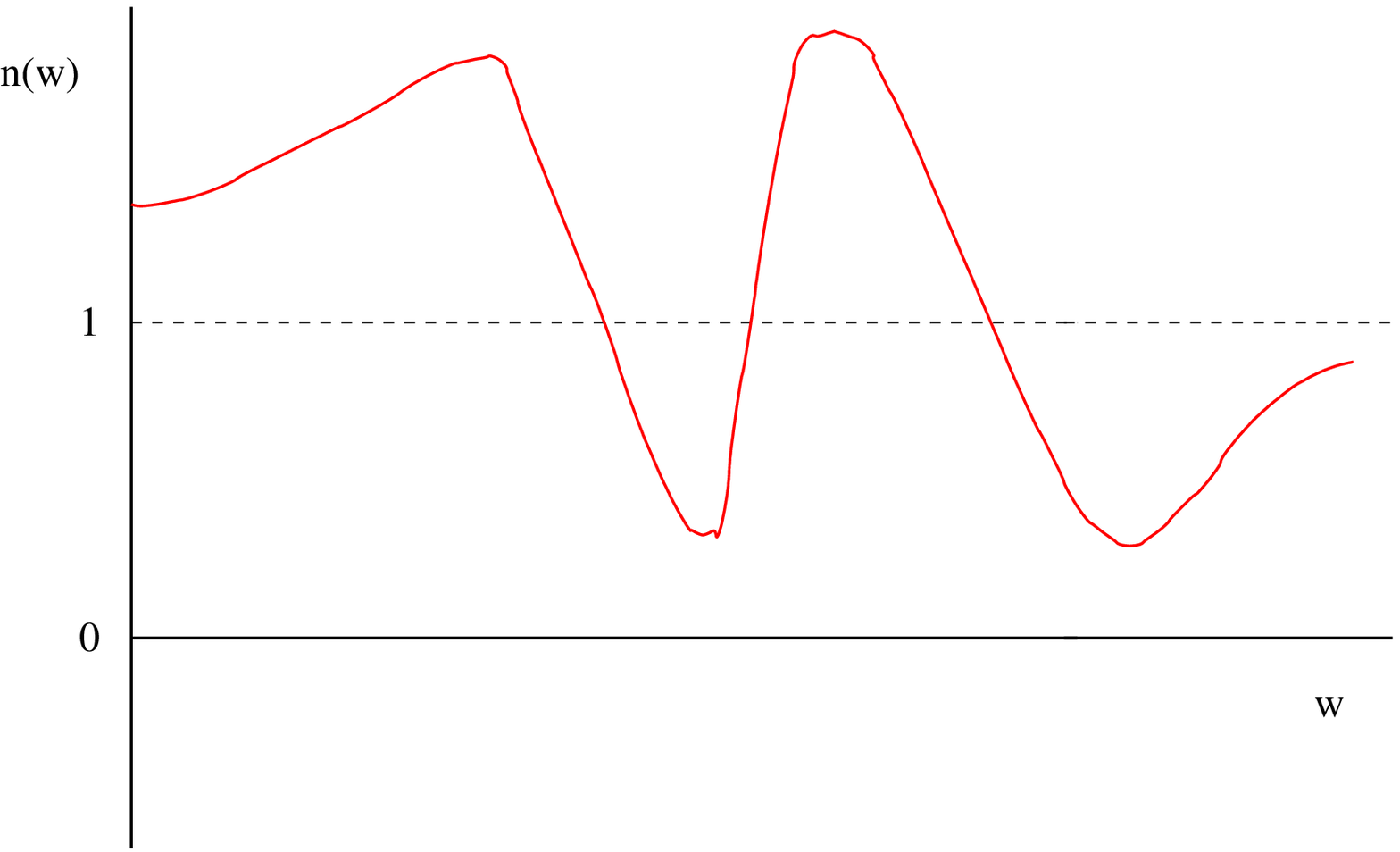}
\caption{The effective refractive index of the cloud of ultra-cold
sodium atoms used in the `slow light' experiments.} \label{fig:5}}
In passing, notice that it is the group velocity which is 
measured in quantum optics experiments which find light speeds of
essentially zero \cite{Hau} or many times $c$ \cite{WKD}. For example, in the
experiments of Vestergaard Hau and colleagues at Harvard, the group velocity 
of a light pulse is reduced almost to zero by shining tuned lasers on a cloud 
of ultra-cold sodium atoms. The set-up is designed to produce an effective
refractive index of the cloud of the form shown in Fig.~5. A very clear explanation
of how this is achieved by manipulating the states of the sodium atoms in
the gas cloud and exploiting quantum superposition is given in ref.\cite{Hau}.
Now, some simple algebra shows that the group velocity can be expressed in terms
of the slope of the refractive index: 
$v_{\rm gp} = \Bigl(n + \w {dn\over d\w}\Bigr)^{-1}$. The group velocity therefore
becomes small in a region where the slope of the refractive index is large.
In the `slow-light' experiments, the central section of the curve $n(\w)$ in Fig.~5
is made as steep as possible and the frequency of the probe laser tuned to this 
frequency range, forcing the group velocity towards zero.

Returning to Fig.~4, it is clear that the phase velocity itself also does not
represent a `speed of light' relevant for considerations of signal
propagation or causality.
The appropriate velocity to define light cones and causality is in fact
the wavefront velocity $v_{\rm wf}$. (Notice that in Fig.~4, $v_{\rm wf}$ is a 
constant, equal to $c$, independent of the frequency or details of the absorption
band.) This is determined by the boundary between the regions of zero
and non-zero disturbance (more generally, a discontinuity in the first
or higher derivative of the field) as the pulse propagates.
Mathematically, this definition of wavefront is identified with the 
characteristics of the partial differential equation governing the 
wave propagation \cite{CH}. Our problem is therefore to determine the velocity
associated with the characteristics of the wave operator derived from the
modified Maxwell equations of motion appropriate to the new effective 
action. 

Notice that a very complete and rigorous discussion of the wave equation in curved
spacetime has already been given in the monograph by Friedlander\cite{Fried}, in 
which it is proved (Theorem 3.2.1) that the characteristics are simply the 
null hypersurfaces of the spacetime manifold, in other words that the 
wavefront always propagates with the fundamental speed $c$. However, this 
discussion assumes the standard form of the (gauge-fixed) Maxwell wave 
equation and does {\it not} cover the modified wave equation (\ref{eq:da}), 
precisely because of the extra curvature couplings which lead to the 
effective metric $G_{\m\n}$ and superluminal propagation.

\subsection{Characteristics and Wavefronts}

Instead, the key result which allows a derivation
of the wavefront velocity is derived by Leontovich \cite{Leon}.
In this paper\footnote{I am very grateful to A. Dolgov, V. Khoze and
I. Khriplovich for their help in obtaining and interpreting ref.\cite{Leon}.}, 
an elegant proof is presented for a very general set of 
PDEs that the wavefront velocity associated with the characteristics is 
identical to the $\w\rta\infty$ limit of the phase velocity, i.e.
\begin{equation}
v_{\rm wf} = \lim_{\w\rta \infty}{\w\over|\ul k|} = 
\lim_{\w\rta \infty}v_{\rm ph}(\w)
\label{eq:fb}
\end{equation}
The proof is rather formal, but is of sufficient generality to apply to our
discussion of photon propagation using the modified effective action
of section 3. We reproduce the essential details below.

The first step is to recognise that any second order PDE can be written as
a system of first order PDEs by considering the first derivatives of the field
as independent variables. Thus, if for simplicity we consider a general second 
order wave equation for a field $u(t,x)$ in one space dimension, the system of
PDEs we need to solve is
\begin{equation}
a_{ij} {\partial\phi_j\over\partial t} + b_{ij} {\partial\phi_j\over\partial x}
+ c_{ij} \phi_j ~=~0
\label{eq:fc}
\end{equation}
where $\phi_i = \{u, {\partial u\over\partial t}, {\partial u\over\partial x}\}$.

Making the `geometric optics' ansatz
\begin{equation}
\phi_i ~=~ \varphi_i \exp i(wt-kx)
\label{eq:fd}
\end{equation}
where the frequency-dependent phase velocity is $v_{\rm ph}(k) = \w(k)/k$, 
and substituting into Eq.(\ref{eq:fc}) we find
\begin{equation}
\Bigl(i\w a_{ij} - ik b_{ij} + c_{ij} \Bigr) \varphi_j ~=~ 0
\label{eq:fe}
\end{equation}
The condition for a solution,
\begin{equation}
{\rm det}\Bigl[a_{ij} v_{\rm ph}(k) - b_{ij} -{i\over k} c_{ij} \Bigr] ~=~ 0
\label{eq:ff}
\end{equation}
then determines the phase velocity.

On the other hand, we need to find the characteristics of Eq.(\ref{eq:fc}),
i.e.~curves ${\cal C}$ on which Cauchy's theorem breaks down and the evolution is not
uniquely determined by the initial data on ${\cal C}$. The derivatives of the field
may be discontinuous across the characteristics and these curves are associated with 
the wavefronts for the propagation of a sharp-fronted pulse. The corresponding light rays
are the `bicharacteristics'. (See, for example, ref.\cite{CH} chapters 5.1, 6.1 for
further discussion.) 

We therefore consider a characteristic curve ${\cal C}$ in the $(t,x)$ plane separating
regions where $\phi_i =0$ (ahead of the wavefront) from $\phi_i \ne 0$ (behind the
wavefront). At a fixed point $(t_0,x_0)$ on ${\cal C}$, the absolute derivative of 
$\phi_i$ along the curve, parametrised as $x(t)$, is just
\begin{equation}
{d\phi_i\over dt} ~=~ {\partial \phi_i\over\partial t}\Big|_0
+ {\partial \phi_i\over\partial x}\Big|_0 {dx\over dt}
\label{eq:fg}
\end{equation}
where $dx/dt = v_{\rm wf}$ gives the wavefront velocity. 
Using this to eliminate ${\partial\phi_i\over\partial t}$ from the PDE Eq.(\ref{eq:fc})
at $(t_0,x_0)$, we find
\begin{equation}
\Bigl( - a_{ij} {dx\over dt} + b_{ij}\Bigr) {\partial \phi_j\over\partial x}\Big|_0
+ a_{ij} {d\phi_j^{(0)}\over dt} + c_{ij}\phi_j^{(0)} ~=~ 0
\label{eq:fh}
\end{equation}
Now since ${\cal C}$ is a wavefront, on one side of which $\phi_i$ vanishes
identically, the second two terms above must be zero.
The condition for the remaining equation to have a solution is simply
\begin{equation}
{\rm det}\Bigl[ a_{ij} v_{\rm wf} - b_{ij} \Bigr] ~=~ 0
\label{eq:fi}
\end{equation}
which determines the wavefront velocity $v_{\rm wf}$.
The proof is now evident. Comparing Eqs.(\ref{eq:ff}) and (\ref{eq:fi}), we
clearly identify
\begin{equation}
v_{\rm wf} ~=~ v_{\rm ph}(k\rightarrow \infty)
\label{eq:fj}
\end{equation}

The wavefront velocity in a gravitational background is therefore
not given {\it a priori} by $c$. Taking vacuum polarisation into account,
there is no simple non-dispersive medium corresponding to the 
vacuum of classical Maxwell theory in which the phase velocity 
represents a true speed of propagation; for QED in curved spacetime, even
the vacuum is dispersive.
In order to discuss causality, we therefore have to extend the original
Drummond-Hathrell results for $v_{\rm ph}(\w \sim 0)$ to the high frequency
limit $v_{\rm ph}(\w\rta\infty)$, as already emphasised in their original 
work. 

A further subtle question arises if we write the standard dispersion relation
for the refractive index $n(\w)$ in the limit $\w\rta\infty$:
\begin{equation}
n(\infty) = n(0) - {2\over\pi} \int_0^\infty {d\w\over\w} \Im n(\w)
\label{eq:fk}
\end{equation}
For a conventional dispersive medium, $\Im n(\w) > 0$, which implies that
$n(\infty) < n(0)$, or equivalently $v_{\rm ph}(\infty) > v_{\rm ph}(0)$.
Evidently this is satisfied by Fig.~4. The key question though is 
whether the usual assumption of positivity of $\Im n(\w)$ holds 
in the present situation of the QED vacuum in a gravitational field.
If so, then (as already noted in ref.\cite{DH}) the superluminal
Drummond-Hathrell results for $v_{\rm ph}(0)$ would actually be {\it lower
bounds} on the all-important wavefront velocity $v_{\rm ph}(\infty)$.
However, it is not clear that positivity of $\Im n(\w)$ holds
in the gravitational context. Indeed it has been explicitly
criticised by Dolgov and Khriplovich in refs.\cite{DK,Khrip},
who point out that since gravity is an inhomogeneous medium in which beam
focusing as well as diverging can happen, a growth 
in amplitude corresponding to $\Im n(\w) < 0$  is possible. 
The possibility of $v_{\rm ph}(\infty) < v_{\rm ph}(0)$, and in particular
$v_{\rm ph}(\infty) = c$, therefore does not seem to be convincingly ruled out
by the dispersion relation Eq.({\ref{eq:fk}).

\subsection{Light in Magnetic Fields}

Before confronting the gravitational case, we can build up some intuition
on dispersion by looking at a closely related situation -- the propagation of
light through a background magnetic field. In this case, the analogue of the
low-frequency, weak-field Drummond-Hathrell action for gravity is the Euler-Heisenberg
action:
\begin{equation}
\C ~=~\int dx \biggl[ -{1\over4} F_{\m\n}F^{\m\n} + {1\over m^4}\biggl(
p F_{\m\n}F_{\l\r}F^{\m\l}F^{\n\r} + q\Bigl( F_{\m\n}F^{\m\n}\Bigr)^2 \biggr)\biggr]
\label{eq:fl}
\end{equation}
where $p = {7\over90}\a^2$ and $q = -{1\over36}\a^2$.
For a constant magnetic field, a familiar geometric optics analysis
produces the following results for the phase velocity of photons 
moving transverse to the background field:
\begin{equation}
k^2 + {8\over m^2}\Bigl[(2p + 4q)_\parallel , p_\perp \Bigr] T_{\m\n}k^\m k^\n ~=~0
\label{eq:fm}
\end{equation}
where $T_{\m\n} = F_\m{}^\l F_{\l\n} - {1\over4}g_{\m\n}F_{\l\r}F^{\l\r}$
is the electromagnetic energy-momentum tensor and the suffices $\parallel$,$\perp$
indicate the two polarisations. This should be compared with Eq.(\ref{eq:di})
for the gravitational case. Notice that for electromagnetism, birefringence appears
already in the energy-momentum tensor term, essentially because the lowest order 
terms in the Euler-Heisenberg action contributing to (\ref{eq:fm}) are of 4th order 
in the background fields, i.e. $O(F^4)$, compared to the 3rd order terms 
of $O(RFF)$ in the Drummond-Hathrell action. Substituting for $p$,$q$ we find the
well-known results for the low-frequency limit of the phase velocity:
\begin{equation}
v_{\rm ph}(\w\rta 0) ~ \sim ~ 
1 - {\a\over4\pi} \Bigl({eB\over m}\Bigr)^2~
\Bigl[{14\over45}_\parallel, {8\over45}_\perp \Bigr]
\label{eq:fn}
\end{equation}
As expected, $v_{\rm ph} < 1$ for both polarisations.

The refractive index for photons of arbitrary frequency has been calculated 
explicitly by Tsai and Erber \cite{TEone,TEtwo}.
They derive an effective action
\begin{equation}
\C = -\int dx \biggl[{1\over4}F_{\m\n} F^{\m\n} ~~
+{1\over2} A^\m M_{\m\n}A^\n \biggr]
\label{eq:fo}
\end{equation}
where $M_{\m\n}$ is a differential operator acting on the electromagnetic
field $A^\n$. Writing the corresponding equation of motion and making
the geometric optics ansatz as usual, we find the light cone condition
\begin{equation}
k^2 -  M_{\m\n}(k) a^\m a^\n ~=~0
\label{eq:fp}
\end{equation}

Denoting $M_{\m\n}a^\m a^\n$ by $M_\parallel, M_\perp$ for the two
polarisations, the complete expression for the birefringent refractive
index is:
\begin{eqnarray}
v_{\rm ph~\parallel,\perp}(\w) ~~&=&~~ 1 ~+~ {1\over2\w^2}~M_{\parallel,\perp} 
\nonumber\\
&=&~~1~+~ 
{\a\over4\pi} \Bigl({eB\over m^2}\Bigr)^2 ~
\int_{-1}^1 du \int_0^\infty ds ~s ~
N_{\parallel,\perp}(u,z)~ e^{-is\bigl(1 + s^2 \W^2 P(u,z)\bigr)} 
\label{eq:fq} 
\end{eqnarray}
where $z = {eB\over m^2}s$ and $\W = {eB\over m^2}{\w\over m}$. 
The functions $N$ and $P$ are given by:
\begin{equation}
P(u,z) = {1\over z^2} \Bigl({\cos zu - \cos z\over 2z\sin z} 
- {1-u^2\over4}\Bigr)
~~=~~{1\over12}(1-u^2) + O(z^2)
\label{eq:fr}
\end{equation}
and
\begin{eqnarray}
N_{\parallel} &= -{\cot z\over2z}\Bigl(1-u^2 + {u\sin zu\over\sin z}\Bigr)
+{\cos zu\over 2z\sin z}~~
=~~ {1\over4} (1-u^2)(1-{1\over3}u^2) + O(z^2) \nonumber\\
{}&{}\nonumber\\
N_{\perp} &= -{z\cos zu\over 2\sin z} +{zu\cot z \sin zu\over 2\sin z} +
{z(\cos zu -\cos z)\over\sin^2 z}~~
=~~ {1\over8} (1-u^2)(1+{1\over3}u^2) + O(z^2) 
\label{eq:fs} 
\end{eqnarray}

In the weak field, low frequency limit, we can disregard the function $P$
and consider only the lowest term in the expansion of $N$ in powers of $z$.
This reproduces the results for $v_{\rm ph}(\w\rta 0)$ shown in Eq.(\ref{eq:fn}).
The weak field, high frequency limit is analysed in ref.\cite{TEtwo}.
It is shown that 
\begin{equation}
v_{\rm ph~\parallel,\perp}(\w\rta \infty) ~ \sim ~
1 + {\a\over4\pi} \Bigl({eB\over m}\Bigr)^2~
\bigl[{c}_\parallel, {c}_\perp \bigr]~ \W^{-{4\over3}}
\label{eq:ft}
\end{equation}
where the numerical constants are $\bigl[{c}_\parallel, {c}_\perp \bigr]
= {3^4 3^{1\over3}\over7}\sqrt{\pi} \C({2\over3})^2 \bigl(\C({1\over6})
\bigr)^{-1}~[3_\parallel,2_\perp]$.

The complete frequency dependence of $v_{\rm ph}(\w)$, or equivalently $n(\w)$,
is therefore known and shows exactly the
features found in the simple absorption model described above.
In particular, the phase velocity $v_{\rm ph}(\w)$ begins less than 1
at low frequencies, showing birefringence but conventional subluminal
behaviour. In the high frequency limit, however, the phase velocity
approaches $c$ from the superluminal side with a $\w^{-{4\over3}}$ 
behaviour. 

The origin of this non-analytic high-frequency behaviour is of particular
interest. It arises from the phase factor $\exp{\bigl[-is^3 \Omega^2 P(u,z)\bigr]}$
in the effective action. The absorption resonance region in the graph of 
the refractive index $n(\w)$ occurs for frequencies $\w$ such that $\Omega \sim 1$.
For high frequencies, even at weak fields, we need $\Omega \rta \infty$.
Clearly, these aspects of the behaviour of $n(\w)$ are invisible to a
simple expansion of the effective action in low powers of the field strength $B$,
even keeping terms of all orders in derivatives. The conclusion is that
a complete analysis of resonant and high frequency propagation requires the
evaluation of the non-perturbative (in a field strength expansion) phase
factor in Eq.(\ref{eq:fq}). As we now see, this poses severe problems for the
analysis of high frequency propagation in the case of a background 
gravitational field.

\subsection{Dispersion and Gravity}

Setting aside this concern for the moment, we now return to gravity and begin 
our study of dispersion by constructing the generalisation of the 
Drummond-Hathrell effective action containing all orders in derivatives,
while retaining only terms of $O(RFF)$ in the curvature and field strength.
This action was recently derived in ref.\cite{Ssix}. The result is:
\begin{eqnarray}
&\nonumber\\
&\C  = \int dx \sqrt{-g} \biggl[-{1\over4}F_{\m\n}F^{\m\n}~
+~{1\over m^2}\Bigl(D_\m F^{\m\l}~ \overrightarrow{G_0}~ D_\n F^\n{}_{\l} \nonumber\\
&~~~~~~+~\overrightarrow{G_1}~ R F_{\m\n} F^{\m\n}~ 
+~\overrightarrow{G_2}~ R_{\m\n} F^{\m\l}F^\n{}_{\l}~
+~\overrightarrow{G_3}~ R_{\m\n\l\r}F^{\m\n}F^{\l\r} \Bigr) \nonumber\\
&~+~{1\over m^4}\Bigl(\overrightarrow{G_4}~ R D_\m F^{\m\l} D_\n F^\n{}_{\l}~ 
+~\overrightarrow{G_9}~ R_{\m\n\l\r} D_\s F^{\s\r}D^\l F^{\m\n} \nonumber\\
&~~~~+~\overrightarrow{G_5}~ R_{\m\n} D_\l F^{\l\m}D_\r F^{\r\n}~
+~\overrightarrow{G_6}~ R_{\m\n} D^\m F^{\l\r}D^\n F_{\l\r}\nonumber \\
&~~~~~~~~~~+~\overrightarrow{G_7}~ R_{\m\n} D^\m D^\n F^{\l\r} F_{\l\r}~
+~\overrightarrow{G_8}~ R_{\m\n} D^\m D^\l F_{\l\r} F^{\r\n}~\Bigr) ~~\biggr] 
\label{eq:fu}
\end{eqnarray}
This was found by adapting a background field action valid to third order in 
generalised curvatures due to Barvinsky, Gusev, Zhytnikov and Vilkovisky \cite{BGVZone} 
(see also ref.\cite{BGVZtwo}) 
and involves re-expressing their more general result in manifestly local form by an 
appropriate choice of basis operators.

In this formula, the $\overrightarrow{G_n}$ ($n\ge 1$) are form factor functions of 
three operators:
\begin{equation}
\overrightarrow{G_n} \equiv G_n\Bigl({D_{(1)}^2\over m^2}, {D_{(2)}^2\over m^2}, 
{D_{(3)}^2\over m^2}\Bigr)
\label{eq:fv}
\end{equation}
where the first entry ($D_{(1)}^2$) acts on the first following term
(the curvature), etc. $\overrightarrow{G_0}$ is similarly defined as a single variable 
function. These form factors are found using heat kernel methods and are
given by `proper time' integrals of known algebraic functions. Their 
explicit expressions can be found in ref.\cite{Ssix}. 
Evidently, Eq.(\ref{eq:fu}) reduces to the Drummond-Hathrell action if we neglect all 
the higher order derivative terms.

The next step is to derive the equation of motion analogous to Eq.(\ref{eq:da})
from this generalised effective action and to apply geometric optics to find
the corresponding light cone. This requires a very careful analysis of the
relative orders of magnitudes of the various contributions to the equation
of motion arising when the factors of $D^2$ in the form factors act on the 
terms of $O(RF)$. These subtleties are explained in detail in ref.\cite{Sfive}.
The final result for the new effective light cone has the form
\begin{equation}
k^2 ~-~ {8\pi\over m^2} F\Bigl({k.D\over m^2}\Bigr) T_{\m\l} k^\m k^\l ~+~
{1\over m^2} G\Bigl({k.D\over m^2}\Bigr) C_{\m\n\l\r} k^\m k^\l a^\n a^\r ~~=~~0
\label{eq:fw}
\end{equation} 
where $F$ and $G$ are known functions with well-understood asymptotic 
properties \cite{Sfive}. Clearly, for agreement with Eq.(\ref{eq:ch}),
we have $F(0) = 2b+4c$, $G(0) = 8c$.

The novel feature of this new light cone condition is that $F$ and $G$ are
functions of the {\it operator} $k.D$ acting on the Ricci and Riemann
tensors.\footnote{Note that because these corrections are already of $O(\a)$,
we can freely use the usual Maxwell relations $k.D k^\n = 0$ and $k.D a^\n = 0$
in these terms; we need only consider the effect of the operator $k.D$ acting
on $R_{\m\n}$ and $R_{\m\n\l\r}$.} So although the asymptotic behaviour of $F$
and $G$ as functions is known, this information is not really useful unless 
the relevant curvatures are eigenvalues of the operator. On the positive side, 
however, $k.D$ does have a clear geometrical interpretation -- it simply describes 
the variation along a null geodesic with tangent vector $k^\m$. 

The utility of this light cone condition therefore seems to hinge on what we know
about the variations along null geodesics of the Ricci and Riemann
(or Weyl) tensors. Unfortunately, we have been unable to find any results in the 
general relativity literature which are valid in a general spacetime. 
To try to build some intuition, we have therefore again looked at particular cases. 
The most interesting example in this case is photon propagation in the Bondi-Sachs 
metric \cite{Bondi,Sachs} which we recently studied in detail \cite{Sfour}.
  
The Bondi-Sachs metric describes the gravitational radiation from an
isolated source. The metric is
\begin{equation}
ds^2 = -W du^2 - 2 e^{2\b} du dr + r^2 h_{ij}(dx^i - U^i du)
(dx^j - U^j du)
\label{eq:fx}
\end{equation}
where
\begin{equation}
h_{ij}dx^i dx^j = {1\over2}(e^{2\c} + e^{2\d}) d\theta^2
+ 2 \sinh(\c - \d) \sin\theta d\theta d\phi
+ {1\over2}(e^{-2\c} + e^{-2\d}) \sin^2\theta d\phi^2
\label{eq:fy}
\end{equation}
The metric is valid in the vicinity of future null infinity ${\cal I}^+$.
The family of hypersurfaces $u = const$ are null, i.e. $g^{\m\n}
\pl_\m u \pl_\n u = 0$. Their normal vector $\ell_\m$ satisfies
\begin{equation}
\ell_\m = \pl_\m u ~~~~~~~~~~~~~\Rightarrow~~~~~
\ell^2 = 0, ~~~~~~~~\ell^\m D_\mu \ell^\n = 0
\label{eq:fz}
\end{equation}
The curves with tangent vector $\ell^\m$ are therefore
null geodesics; the coordinate $r$ is a radial parameter along these rays  
and is identified as the luminosity distance. 
The six independent functions  $W,\b,\c,\d,U^i$
characterising the metric have expansions in 
${1\over r}$ in the asymptotic region near ${\cal I}^+$, the coefficients of
which describe the various features of the gravitational radiation.

In the low frequency limit, the light cone is given simply by Eq.(\ref{eq:di})
with $T_{\m\n} = 0$. 
The velocity shift is quite different for the case of outgoing
and incoming photons \cite{Sfour}. For outgoing photons, $k^\m = \w \ell^\m$,
and the light cone, written in Newman-Penrose notation\footnote{Details of the 
Newman-Penrose formalism can be found, e.g., in refs.\cite{Sthree,Sfour}.
The essential idea is to construct a basis of null vectors $\ell^\m$, $n^\m$,
$m^\m$ and $\bar m^\m$, and refer the components of the Ricci and Weyl
tensors to this basis. The independent components of the Weyl tensor 
then form a set of 5 complex scalars, denoted $\Psi_0, \ldots, \Psi_4$.
For example, $\Psi_0 = -C_{\m\n\l\r}\ell^\m m^\n \ell^\l m^\r$. In the example
above, we identify the momentum direction for outgoing photons with $\ell^\m$; 
$m^\m$ is then the complex linear combination $a_{\parallel} + ia_{\perp}$
of the physical transverse polarisations.} is
\begin{equation}
k^2 ~~=~~\pm~{4c\w^2\over m^2}~\Bigl(\Psi_0 + \Psi_0^*\Bigr) ~~\sim ~~
O\Bigl({1\over r^5}\Bigr)
\label{eq:fza}
\end{equation}
while for incoming photons, $k^\m = \w n^\m$,
\begin{equation}
k^2 ~~=~~\pm~{4c\w^2\over m^2}~\Bigl(\Psi_4 + \Psi_4^*\Bigr) ~~\sim ~~
O\Bigl({1\over r}\Bigr)
\label{eq:fzb}
\end{equation}

Now, it is a special feature of the Bondi-Sachs spacetime 
that the absolute derivatives of each of the Weyl
scalars $\Psi_0, \ldots, \Psi_4$ along the ray direction $\ell^\m$
vanishes, i.e.~$\Psi_0, \ldots, \Psi_4$ are parallel transported along 
the rays \cite{Sachs,Inverno}. In this case, therefore, we have: 
\begin{equation}
\ell\cdot D ~\Psi_0 ~=~0 ~~~~~~~~~~~~~~~~~~~~~~~~~~~~
\ell\cdot D ~\Psi_4 ~=~0
\label{eq:fzc}
\end{equation}
but there is no equivalent simple result for either $n\cdot D ~\Psi_4$
or $n\cdot D ~\Psi_0$.

Although it is just a special case, Eq.(\ref{eq:fzc}) nevertheless 
leads to a remarkable conclusion. The full light cone condition 
Eq.(\ref{eq:fw}) applied to outgoing photons in the Bondi-Sachs spacetime now 
reduces to
\begin{equation}
k^2 ~\pm~ {\w^2\over 2 m^2}~ G\bigl(0\bigr)
\Bigl(\Psi_0 + \Psi_0^*\Bigr) ~~=~~ 0
\label{eq:fzd}
\end{equation}
since $\ell\cdot D \Psi_0 = 0$. In other words, the low-frequency 
Drummond-Hathrell prediction of a superluminal phase velocity $v_{\rm ph}(0)$
is {\it exact} for all frequencies. There is no dispersion, and the 
wavefront velocity $v_{\rm ph}(\infty)$ is indeed superluminal. 

This is potentially a very important result. It appears to show that
there is at least one example in which the wavefront truly propagates with 
superluminal velocity. If so, quantum effects
would indeed have shifted the light cone into the geometrically spacelike region.

However, we now have to confront the problem raised above in the discussion 
of light propagation in background magnetic fields.
If the gravitational case is similar, this would imply that the modified
light cone can be written heuristically as
\begin{equation}
k^2 ~+~ {\a\over\pi} \int_0^\infty ds~{\cal N}(s,R)~
e^{-is\bigl(1 + s^2\W^2 {\cal P}(s,R)\bigr)}~~=~~0
\label{eq:fze}
\end{equation}
where both ${\cal N}$ and ${\cal P}$ can be expanded in powers of curvature,
and derivatives of curvature, presumably associated with factors of $\w$ as in 
the last section. The frequency
dependent factor $\W$ would be $\W \sim {R\over m^2}{\w\over m} \sim
O({\l_c^3\over\l L^2})$, where `$R$' denotes some generic curvature component 
and $L$ is the typical curvature scale. If this is true, then an expansion of 
the effective action to $O(RFF)$, even including higher derivatives, 
would not be sufficient to reproduce the full, non-perturbative contribution 
$\exp\bigl[-is^3\Omega^2{\cal P}\bigr]$. The Drummond-Hathrell action would correspond 
to the leading order term in the expansion of Eq.(\ref{eq:fze}) in powers of 
${R\over m^2}$ neglecting derivatives, while our improved effective action
(\ref{eq:fu}) sums up all orders in derivatives while retaining the restriction 
to leading order in curvature.  

The omission of the non-perturbative contribution would be justified only in the 
limit of small $\W$, i.e.~for ${\l_c^3\over\l L^2} \ll 1$. Neglecting this 
therefore prevents us from accessing the genuinely high frequency limit
$\l \rta 0$ needed to find the asymptotic limit $v_{\rm ph}(\infty)$ of the phase 
velocity. Moreover, assuming Eq.(\ref{eq:fze}) is indeed on the right lines, 
it also seems inevitable that for high frequencies (large $\W$) the rapid phase
variation in the exponent will drive the entire heat kernel integral to zero,
ensuring the wavefront velocity $v_{\rm wf} = c$.

At present, it is not clear how to make further progress. The quantum field theoretic
calculation required to find such non-perturbative contributions to the effective 
action and confirm an $\exp\bigl[-is^3\Omega^2{\cal P}\bigr]$ structure in 
Eq.(\ref{eq:fze}) appears difficult, although some technical progress in this area 
has been made recently in ref.\cite{BM} and work in progress \cite{Gusev}.
One of the main difficulties is that since a superluminal effect requires 
some anisotropy in the curvature, it is not sufficient just to consider 
constant curvature spacetimes. (Recall that the Ricci scalar term in the 
effective action (\ref{eq:ch}) does not contribute to the modified light 
cone (\ref{eq:db}).) A possible approach to this problem, which would help
to control the plethora of indices associated with the curvatures, might be 
to reformulate the heat kernel calculations directly in the Newman-Penrose
basis. On the other hand, perhaps a less ambitious goal would be to try to
determine just the asymptotic form of the non-perturbative contribution in the 
$\Omega\rta\infty$ limit.

A final resolution of the dispersion problem for QED in curved spacetime has 
therefore still to be found.  At present, the most likely scenario appears to be 
that high-frequency dispersion is driven by non-perturbative contributions to the 
effective action such that the wavefront velocity remains precisely $c$. 
It would then be interesting to see exactly how the phase velocity behaves as a 
function of $\w$. What we have established is that for `sub-resonant' frequencies
the phase velocity may, depending on the polarisation, take both the conventional 
form of Fig.~4 or the superluminal mirror image where $v_{\rm ph}(0) > 1$. 
We have even seen in the Bondi-Sachs example a special case where 
$v_{\rm ph}(\w) \sim {\rm const} > 1$ for relatively low frequencies.

However, even if this picture is correct and the light cone is eventually
driven back to $k^2=0$ in the high frequency limit, the analysis described here
still represents a crucial extension of the domain of validity of
the superluminal velocity prediction of Drummond and Hathrell. Recall
from section 4.3 that the constraint on the frequency for which the 
superluminal effect is in principle observable is ${\l_c^2\over\l L} \gg 1$.
Obviously this was not satisfied by the original $\w\sim 0$ derivation.
However, our extension based on the generalised effective
action (\ref{eq:fu}) does satisfy this constraint. Combining with the restriction
${\l_c^3\over \l L^2} \ll 1$ in which the neglect of the ${\cal P}$ type
corrections is justified, we see that there is a frequency range
\begin{equation}
{\l_c\over L} ~~\gg ~~ {\l\over \l_c} ~~\gg~~ {\l_c^2\over L^2}
\label{eq:fzf}
\end{equation}
below the `resonance' region of Fig.~4, 
for which our expression (\ref{eq:fw})
for the modified light cone is valid and predicts in principle observable effects.

Since this formula allows superluminal corrections to both the phase and the signal
velocities, we conclude that superluminal propagation has indeed been established as
an observable phenomenon even if, as seems likely, causality turns out to 
respected through the restoration of the standard light cone $k^2=0$ 
in the asymptotic high frequency limit.

\section{Speculations}

In this paper, we have reviewed some of the qualitatively new phenomena that occur
when quantum theory is introduced into the already rich field of gravitational 
optics. The picture that arises is of curved spacetime as an optical medium,
characterised by a novel refractive index and displaying polarisation-dependent
properties such as birefringence. 

Naturally, since these phenomena arise due to intrinsically quantum field theoretic
processes such as vacuum polarisation, they are extremely small for the curvatures
typical of present day astrophysics. A typical order of magnitude for the shift
in the speed of light around a black hole, for example, is 
$O\bigl(\a m_{pl}^4/(m^2 M^2)\bigr)$, where $m$ is the electron mass setting the 
scale for vacuum polarisation and $M$ is the black hole mass. Quantum gravitational
optics (QGO) is therefore only likely to play an important role in situations of extreme,
quantum-scale, curvature such as may be present in the very early universe or in the
vicinity of primordial black holes. Its main interest, as in the analogous case of 
Hawking radiation, is likely to be primarily conceptual.

Indeed, serious conceptual issues do arise through the apparent prediction in certain
circumstances of a superluminal speed of light. We have explained carefully how this
occurs and discussed the subtle questions of interpretation that arise. Although
further work remains to be done to achieve a full understanding, it appears that
despite an anomalous refractive index featuring a superluminal low-frequency
phase (and signal) velocity, the critical high-frequency limit which determines the
characteristics or wavefront velocity remains equal to $c$. Nonetheless, we described
how superluminal velocities could be accommodated in the framework of general relativity
while preserving the essential notion of stable causality.

All this work was within the relatively uncontroversial theoretical formalism of
quantum electrodynamics in curved spacetime. To close, we give a brief survey of
other more speculative ideas in the current literature which overlap in some way
with the ideas presented here.

First, we should note that everything that has been said about the QED effective action
arises equally in the low-energy effective actions derived from string theory. 
SEP-violating interactions of $O(RFF)$ naturally occur there too, with the main difference 
that the electron scale $m$ becomes the Planck mass $m_{pl}$. Evidently this simply
reduces the relative order of magnitude by another huge factor. 

One of the key insights in the theory presented here is that in QGO the usual light cone
relation for the photon momentum becomes
\begin{equation}
G_{\m\n} p^\m p^\n  = 0
\label{eq:ga}
\end{equation}
That is, the propagation of real photons is governed by a new metric $G_{\m\n}$,
which may in general depend not only on the curvature and its derivatives but also
on the photon direction, polarisation and energy. The physical light cones are then
distinct from the geometric null cones characterised by the spacetime metric $g_{\m\n}$.
This realisation of QGO is therefore intrinsically a ${\it bimetric}$ theory.

Inspired by this, it is interesting to speculate on modifications to conventional
general relativity in which a bimetric structure is imposed {\it a priori}.
Phenomenologically, this has the advantage that the new scale characterising the 
propagation metric can be a free parameter, with the result that the type of phenomena
discussed here could occur on macroscopically accessible scales. As a result, there
is a strong motivation to search for unusual velocity or polarisation dependent
phenomena in light signals from cosmological sources. Bimetric theories have a long
history, with a particularly elegant recent construction due to Drummond \cite{Drum},
who has proposed an extension of general relativity in which matter couples to its
own vierbein, which is rotated relative to the geometric vierbein according to
an appropriate sigma model dynamics. By modifying gravitational dynamics on an
astrophysical scale, such theories have the potential to offer an alternative to
dark matter models. Other recent work exploiting bimetric theories to resolve
cosmological puzzles can be found in a series of papers on scalar-tensor gravity
by Clayton and Moffat \cite{Moff,Mofftwo}.

These theories modify general relativity, and by incorporating a bimetric structure
they necessarily involve a varying speed of light (VSL). VSL theories and their
potential for providing an alternative to inflation as a resolution of cosmological
problems such as the horizon problem have seen a rapid rise in popularity in recent
years. A small selection of relevant papers is included here as an introduction
to this literature for the interested reader \cite{VSLone,VSLtwo,VSLthree,VSLfour,
VSLfive,VSLsix}.

The QED version of QGO discussed in this paper, and some of the bimetric models,
preserve the essential principles of general relativity, especially the weak
equivalence principle and local Lorentz invariance. We can of course speculate
that local Lorentz invariance itself may be violated. A phenomenological approach
to such theories has been pioneered by Kostelecky, who has constructed a
general extension to the standard model incorporating Lorentz and CPT violating
interactions and investigating their experimental consequences \cite{Kost}. Of particular
interest here are the following interactions, which affect the propagation of
light:
\begin{equation}
\C = \C_{\rm M} + \int dx \biggl[-{1\over4} K_{\m\n\l\r}~F^{\m\n}F^{\l\r} 
~+~{1\over2} L^\m~ \e_{\m\n\l\r} A^\n F^{\l\r} \biggr]
\label{eq:gb}
\end{equation}
Here, $K$ and $L$ are Lorentz-violating couplings, which could arise as VEVs
in some more fundamental theory exhibiting spontaneous Lorentz breaking.
Evidently, the first of these interactions is the analogue of the SEP-violating
interaction considered in this paper, with the replacement of the covariant
curvature tensor $R_{\m\n\l\r}$ by the coupling $K_{\m\n\l\r}$. The phenomenology
of light propagation, and some (though not all) of the considerations concerning
causality discussed here therefore apply directly to this class of Lorentz-violating 
theory. The CPT violating Chern-Simons interaction does not of course arise in QED,
though it can be induced by anomalies involving spacetime topology \cite{Klink}.
The Kostelecky lagrangian has been used as a phenomenological model against
which to put limits on the magnitude of VSL, polarisation-dependent or 
Lorentz-violating effects in light propagation. Since a birefringent speed of
light would cause a rotation of the polarisation of light as it travels through
space, the observation of polarisation from distant quasars and radio galaxies,
and also recent measurements of CMB polarisation, can be used to place limits
on the phenomenological couplings $K$ and $L$. Details of this work can be
found in ref.\cite{CFJ}. Similar interactions are induced also in 
non-commutative gauge theories as a consequence of the Moyal product, with
similar phenomenological implications. Here, the
analogue of the coupling $K_{\m\n\l\r}$ is constructed from the 
non-commutativity parameter $\theta_{\m\n}$ defined from 
$[x^\m,x^\n] = i\theta^{\m\n}$. 

Of course, once we permit models with more or less {\it ad hoc} Lorentz violation,
then we are free to speculate at will as to alternative dispersion relations
for the propagation of light, or indeed neutrinos. Some recent popular
models invoke a supposed non-realisation of Lorentz invariance in the 
effective low-energy field theory derived from quantum gravity 
\cite{QGone,QGtwo,King}, non-linear realisations of Lorentz symmetry preserving the
Planck length as an invariant \cite{Magsmol}, so-called `doubly special'
relativity \cite{Amelino}, etc. Evidently, there is immense freedom to frame
such models within the constraints of experimental data though it is far less
clear how to incorporate them into complete theories consistent with our
understanding of unified theories of particle physics and quantum gravity. 
On the experimental side, particular interest attaches to light (or neutrinos)
received from gamma-ray bursts, since the extremely high energies of the photons
can enhance the size of the VSL effect for certain Lorentz-violating 
dispersion relations. 

This rapid tour of admittedly rather speculative models is hopefully sufficient
to indicate the considerable current interest in looking for anomalies in the
standard theory of light propagation in special and general relativity. 
However, as we have shown in this paper, it is not necessary to invoke
{\it ad hoc} violations of the fundamental principles of quantum field theory,
general relativity or string theory in order to discover a new richness 
of phenomena in the propagation of light. Whether its primary interest will lie
in the conceptual issues it raises for the consistency of quantum field theory
with gravity, or in the prediction of experimental anomalies in the 
speed of light in cosmology, the field of quantum gravitational optics
is promised a bright future.

\acknowledgments

I am especially grateful to I. Drummond for many interesting discussions
on superluminal propagation and numerous other colleagues,
including A. Dolgov, G. Gibbons, H. Gies, V. Khoze, S. King, K. Kunze, M. Maggiore, 
J. Magueijo, H.Osborn, W. Perkins, S. Sarkar and G. Veneziano, for their comments 
on this work.
This research is supported in part by PPARC grant PP/G/O/2000/00448.


\begin{thebibliography}{999}

\bibitem{DH}{I.T. Drummond and S. Hathrell, Phys. Rev. D22 (1980) 343. }
\bibitem{Hawkrad}{S.W. Hawking, Commun. Math. Phys. 43 (1975) 199. }
\bibitem{Sfive}{G.M. Shore, Nucl. Phys. B633 (2002) 271.}
\bibitem{Ssix}{G.M. Shore, Nucl. Phys. B646 (2002) 281.}
\bibitem{Seight}{G.M. Shore, {\it Superluminal Light}, to appear in the Proceedings:
`Time and Matter: An International Colloquium on the Science of Time', Venice 2002,
eds. I. Bigi and M. Faessler, gr-qc/0302116.} 
\bibitem{SEF}{P. Schneider, J. Ehlers and E.E. Falco, {\it Gravitational Lenses},
Springer-Verlag, Berlin, 1992.}
\bibitem{Sthree}{G.M. Shore, Nucl. Phys. B460 (1996) 379. }
\bibitem{Scharn}{K. Scharnhorst, Phys. Lett. B236 (1990) 354.}
\bibitem{Barton}{G. Barton, Phys. Lett. B237 (1990) 559. } 
\bibitem{LPT}{J. I. Latorre, P. Pascual and R. Tarrach, Nucl. Phys. B437 (1995) 60.}
\bibitem{Gies}{W. Dittrich and H. Gies, Phys. Lett. B431 (1998) 420-429;
Phys. Rev. D58 (1998) 025004.}
\bibitem{Sone}{R.D. Daniels and G.M. Shore, Nucl. Phys. B425 (1994) 634. }
\bibitem{Stwo}{R.D. Daniels and G.M. Shore, Phys. Lett. B367 (1996) 75. }
\bibitem{Gibb}{G.W. Gibbons and C.A.R. Herdeiro, Phys. Rev. D63 (2001) 064006.}
\bibitem{Hawk}{S.W. Hawking, {\it The Event Horizon}, 1972 Les Houches lectures,
ed. B. De Witt, Gordon and Breach, 1972.}
\bibitem{HE}{S.W. Hawking and G.F.R. Ellis, {\it The Large Scale Structure of
Spacetime}, Cambridge University Press, 1973.}
\bibitem{LSV}{S. Liberati, S. Sonego and M. Visser, Annals Phys. 298 (2002) 167.}
\bibitem{Brill}{L. Brillouin, {\it Wave Propagation and Group Velocity}, 
Academic Press, London, 1960.}
\bibitem{Hau}{L.V. Hau, Scientific American 7 (2001) 66.}
\bibitem{WKD}{L.J. Wang, A. Kuzmich and A. Dogoriu, Nature 406 (2000) 277.}
\bibitem{CH}{R. Courant and D. Hilbert, {\it Methods of Mathematical Physics, Vol II},
Interscience, New York, 1962.}
\bibitem{Fried}{F.G. Friedlander, {\it The Wave Equation on a Curved Spacetime},
Cambridge University Press, 1975.}
\bibitem{Leon}{M.A. Leontovich, {\it in} L.I. Mandelshtam,
{\it Lectures in Optics, Relativity and Quantum Mechanics}, Nauka, Moscow 1972~
{(\it in Russian).}}
\bibitem{DK}{A.D. Dolgov and I.B. Khriplovich, Sov. Phys. JETP 58(4) (1983) 671.}
\bibitem{Khrip}{I.B. Khriplovich, Phys. Lett. B346 (1995) 251.}
\bibitem{BGVZone}{A.O. Barvinsky, Yu.V. Gusev, G.A. Vilkovisky and V.V. Zhytnikov,
Print-93-0274 (Manitoba), 1993.}
\bibitem{BGVZtwo}{A.O. Barvinsky, Yu.V. Gusev, G.A. Vilkovisky and V.V. Zhytnikov,
J. Math. Phys. 35 (1994) 3525; J. Math. Phys. 35 (1994) 3543; 
Nucl. Phys. B439 (1995) 561.}
\bibitem{Chand}{S. Chandresekhar, {\it The Mathematical Theory of Black Holes},
Clarendon, Oxford, 1985.}
\bibitem{Bondi}{H. Bondi, M.G.J. van der Burg and A.W.K. Metzner, Proc. Roy. Soc.
A269 (1962) 21}
\bibitem{Sachs}{R.K. Sachs, Proc. Roy. Soc. A270 (1962) 103.}
\bibitem{Sfour}{G.M. Shore, Nucl. Phys. B605 (2001) 455. }
\bibitem{Inverno}{R.A. d'Inverno, {\it Introducing Einstein's Relativity},
Clarendon, Oxford, 1992.}
\bibitem{TEone}{W. Tsai and T. Erber, Phys. Rev. D10 (1974) 492.}
\bibitem{TEtwo}{W. Tsai and T. Erber, Phys. Rev. D12 (1975) 1132.}
\bibitem{Adler}{S. Adler, Ann. Phys. (N.Y.) 67 (1971) 599.}
\bibitem{BM}{A.O. Barvinsky and V.F. Mukhanov, Phys. Rev. D66 (2002) 065007.} 
\bibitem{Gusev}{Yu.V. Gusev, private communication.}
\bibitem{Drum}{I.T. Drummond, Phys. Rev. D63 (2001) 043503.}
\bibitem{Moff}{M.A. Clayton and J.W. Moffat, Phys. Lett. B506 (2001) 177; astro-ph/0203164.}
\bibitem{Mofftwo}{J.W. Moffat, Int. J. Mod. Phys. D12 (2003) 281.} 
\bibitem{VSLone}{J.W. Moffatt, Int. J. Mod. Phys. D2 (1993) 351.} 
\bibitem{VSLtwo}{A. Albrecht and J. Magueijo, Phys. Rev. D59 (1999) 043516.}
\bibitem{VSLthree}{J.D. Barrow, Phys. Rev. D59 (1999) 043515.}
\bibitem{VSLfour}{J.D. Barrow and J. Magueijo, Phys. Lett. B447 (1999) 246.}
\bibitem{VSLfive}{M.A. Clayton and J.W. Moffatt, Phys. Lett. B460 (1999) 263; 
Phys. Lett. B447 (2000) 269.}
\bibitem{VSLsix}{B.A. Bassett, S. Liberati, C. Molina-Paris and M. Visser, 
Phys. Rev. D62 (2000) 103518.}
\bibitem{Kost}{D. Colladay and V.A. Kostelecky, Phys. Rev. D58 (1998) 116002.}
\bibitem{Klink}{F.R. Klinkhamer,  {\it CPT Violation: Mechanism and Phenomenology},
to appear in the Proceedings: `Seventh International Wigner Symposium',
ed. M.E. Noz, hep-th/0110135.}
\bibitem{CFJ}{S.M. Carroll, G.B. Field and R. Jackiw, Phys. Rev. D41 (1990) 1231.}
\bibitem{QGone}{G. Amelino-Camelia, J.R. Ellis, N.E. Mavramatos, D.V. Nanopoulos and S. Sarkar,
Nature 393 (1998) 763.}
\bibitem{QGtwo}{J. Alfaro, H.A. Morales-Tecotl and L.F. Urrutia, Phys. Rev. Lett. 84
(2000) 2318.}
\bibitem{King}{S. Choubey and S.F. King, hep-ph/0207260.} 
\bibitem{Magsmol}{J. Magueijo and L. Smolin, Phys. Rev. D67 (2003) 044017.}
\bibitem{Amelino}{G. Amelino-Camelia, Int. J. Mod. Phys. D11 (2002) 1643.}





\end{thebibliography}
\end{document}